\documentclass[12pt, TexShade, letterpaper]{article}

\usepackage{tabularx} 
\usepackage{amsmath}  
\usepackage[margin=1in,letterpaper]{geometry} 
\usepackage{cite} 
\usepackage[final]{hyperref} 
\usepackage{authblk}
\usepackage{colortbl}

\newcounter{num}

\usepackage[pdftex]{graphicx}
\usepackage[labelfont=bf]{caption}
\usepackage{floatrow} 
\floatstyle{plaintop}
\newcommand{\Figref}[1]{Figure \ref{#1}}

\usepackage{tabularx} 
\usepackage{tabu}
\restylefloat{table}
\newcommand{\Tabref}[1]{Table \ref{#1}}

\usepackage{amsmath}
\usepackage{amssymb}
\usepackage{mathtools}

\newcommand{\Equref}[1]{Equation \ref{#1}}

\usepackage{cite}
\usepackage{url}
\usepackage[nottoc]{tocbibind}
\usepackage{url}

\usepackage{color}
\usepackage{soul}

\usepackage{tikz}
\usepackage{tikz-feynhand}
\usetikzlibrary{intersections,calc,positioning,shadows.blur,decorations.pathreplacing}
\usepackage{etoolbox}

\tikzset{%
        brace/.style = { decorate, decoration={brace, amplitude=5pt} },
       mbrace/.style = { decorate, decoration={brace, amplitude=5pt, mirror} },
        label/.style = { black, midway, scale=0.5, align=center },
     toplabel/.style = { label, above=.5em, anchor=south },
    leftlabel/.style = { label,rotate=-90,left=.5em,anchor=north },   
  bottomlabel/.style = { label, below=.5em, anchor=north },
        force/.style = { rotate=-90,scale=0.4 },
        round/.style = { rounded corners=2mm },
       legend/.style = { right,scale=0.4 },
        nosep/.style = { inner sep=0pt },
   generation/.style = { anchor=base }
}

\tikzset{
    photon/.style={decorate, decoration={snake}, draw=black},
    electron/.style={draw=black, postaction={decorate},
        decoration={markings,mark=at position .55 with {\arrow[draw=black]{>}}}},
    positron/.style={draw=black, postaction={decorate},
        decoration={markings,mark=at position .55 with {\arrow[draw=black]{<}}}},
    gluon/.style={decorate, draw=black,
        decoration={coil,amplitude=4pt, segment length=5pt}},
}

\usetikzlibrary{arrows.meta}
\tikzset{%
  >={Latex[width=2mm,length=2mm]},
            base/.style = {rectangle, draw=black,
                           minimum width=4cm, minimum height=1cm,
                           text centered, font=\sffamily},
             base2/.style = {rectangle, rounded corners,draw=black,
                           minimum width=4cm, minimum height=0.7cm,
                           text centered, font=\sffamily},
  blue/.style = {base, fill=blue!30, minimum height=0.7cm},
       startstop/.style = {base, fill=red!30},
    activityRuns/.style = {base, fill=green!30},
         process/.style = {base2, minimum width=2.5cm, fill=orange!15,},
        }
                          
\usetikzlibrary{trees}
\usetikzlibrary{decorations.pathmorphing}
\usetikzlibrary{decorations.markings}

\begin{document}

\title{Study of $e^+e^- \to \gamma h$ at the ILC}

\author{Yumi Aoki$^{1}$\footnote{yumia@post.kek.jp}, Keisuke Fujii$^{2}$\footnote{keisuke.fujii@kek.jp}, Junping Tian$^{3}$\footnote{tian@icepp.s.u-tokyo.ac.jp}\\
on behalf of the ILD concept group}

\affil{SOKENDAI$^{1}$, KEK$^{2}$, University of Tokyo$^{3}$}
\date{}

\maketitle

\begin{abstract}
We studied the $e^+e^- \to h \gamma$ process at the full simulation level, using a realistic detector model to study the feasibility to constrain the SM effective field theory (SMEFT) $h \gamma Z$ coefficient, $\zeta_{AZ}$, at the ILC. Assuming International Large Detector (ILD) operating at 250 GeV ILC, it is shown that the $e^+e^- \to h\gamma$ process is much more difficult to observe than naively expected if there is no BSM contribution. We thus put upper limits on the cross section of this process. The expected combined 95\% C.L. upper limits for full polarisations $(P_{e^-}, P_{e^+})=(-100\%, +100\%)$ and $(+100\%, -100\%)$ are $\frac{\sigma_{h \gamma}^{L}}{\sigma_{S M}^{L}}<5.0$ and $\frac{\sigma_{h \gamma}^{R}}{\sigma_{S M}^{R}}<61.9$, respectively. The resultant 95\% C.L. limit on $\zeta_{AZ}$ is $-0.020<\zeta_{A Z}<0.003$.
\footnote{
This study has been performed in the framework of the ILD concept group.}

\end{abstract}

\section{Introduction}
In the SM effective field theory (SMEFT) Lagrangian there are ten CP-even operators relevant for Higgs coupling measurements. Among the ten, we focus on those regarding $h \gamma \gamma$ and $h \gamma Z$ couplings and study how the $e^+e^- \to \gamma h$ process constrain them. 

In the SMEFT Lagrangian, the terms relevant to the $h \gamma \gamma$ and $h \gamma Z$ couplings are
\begin{eqnarray}
{\Delta \mathcal{L}_{h\gamma \gamma, h \gamma Z}  } = \mathcal{L}_{SM} + \frac { \zeta _ { A Z } } { v } A _ { \mu \nu } Z ^ { \mu \nu } h + \frac { \zeta _ { A } } { 2 v } A _ { \mu \nu } A ^ { \mu \nu } h,
\label{lsm}
\end{eqnarray}
with
\begin{eqnarray}
\zeta_{A}&=&8 s_{w}^{2}\left(\left(8 c_{W W}\right)-2\left(8 c_{W B}\right)+\left(8 c_{B B}\right)\right)\\
\zeta_{A Z}&=&s_{w} c_{w}\left(\left(8 c_{W W}\right)-\left(1-\frac{s_{w}^{2}}{c_{w}^{2}}\right)\left(8 c_{W B}\right)-\frac{s_{w}^{2}}{c_{w}^{2}}\left(8 c_{B B}\right)\right),
\end{eqnarray}
where $A_{\mu \nu}$ and $Z_{\mu \nu}$ are field strength tensors for photon and $Z$ boson, $v$ is the Higgs vacuum expectation value, $s_w \equiv \sin_{\theta_{W}}$, $c_w \equiv \cos_{\theta_{W}}$, $c_{W W}$, $c_{W B}$, and $c_{B B}$ are SMEFT coefficient in the Warsaw basis~\cite{2018}.

According to HL-LHC projection~\cite{ATLAS:2018jlh}, the measurement of the $h \to \gamma \gamma$ branching ratio can constrain the anomalous $h \gamma \gamma$ coupling, $\zeta_{A}$, rather strongly to 4\%. On the other hand, since the expected precision of the $h \to \gamma Z$ branching ratio is much worse, about 20\% level for the statistical uncertainties, experimental systematic uncertainties, and theory uncertainties in the modeling of the signal and background processes~\cite{ATLAS:2018jlh}, the expected constraint on the anomalous $h \gamma Z$ coupling, $\zeta_{AZ}$, will be much weaker. It is hence important to study how the measurement of the $e^+e^- \to \gamma h$ process constrain the $h\gamma Z$ coupling.
%

%
At the ILC, the Higgs boson can be produced in association with a photon ($e^+e^- \to \gamma h$). Previous study, which is based on a fast simulation and simple cut-based event selection,~\cite{qing2015probing} showed that the $e^+e^- \to h \gamma$ process is difficult to observe in the SM case, but can be used to discover various BSM scenarios.
We therefore carried out full simulation study using a realistic detailed model of ILD and essentially all of potentially relevant SM backgrounds.

\section{Theoretical Framework}
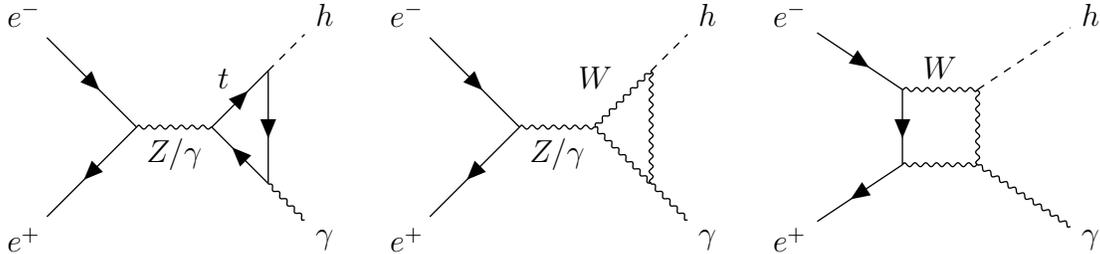
\begin{figure}[htbp] 
        \centering
	\begin{minipage}{0.3\hsize}
	\begin{flushleft}
	\begin{tikzpicture} 
\begin{feynhand}    
    \vertex [particle] (i1) at (-2,-1.5) {$e^+$};
    \vertex [particle] (i2) at (-2,1.5) {$e^-$};
    \vertex [particle] (f1) at (2,-1.5) {$\gamma$};
    \vertex [particle] (f2) at (2,1.5) {$h$};
    \vertex [particle] (l2) at (1.25,0.75);
    \vertex [particle] (l1) at (1.25,-0.75);

    \vertex (w1) at (-0.5,0);
    \vertex (w2) at (0.5,0);
    
    \propag [fermion] (w1) to (i1);
    \propag [fermion] (i2) to (w1);
    
    \propag [fermion] (w2) to [edge label=$t$] (l2);
    \propag [fermion] (l2) to (l1);
    \propag [fermion] (l1) to (w2);
    
    \propag [scalar] (l2) to (f2);
    \propag [boson] (l1) to (f1);

    \propag [boson] (w2) to [edge label=$Z/\gamma$] (w1);
\end{feynhand}
\end{tikzpicture}
	\end{flushleft}
	\end{minipage}
	\begin{minipage}{0.3\hsize}
	\begin{flushleft}
		\begin{tikzpicture} 
\begin{feynhand}    
    \vertex [particle] (i1) at (-2,-1.5) {$e^+$};
    \vertex [particle] (i2) at (-2,1.5) {$e^-$};
    \vertex [particle] (f1) at (2,-1.5) {$\gamma$};
    \vertex [particle] (f2) at (2,1.5) {$h$};
    \vertex [particle] (l2) at (1.25,0.75);
    \vertex [particle] (l1) at (1.25,-0.75);

    \vertex (w1) at (-0.5,0);
    \vertex (w2) at (0.5,0);
    
    \propag [fermion] (w1) to (i1);
    \propag [fermion] (i2) to (w1);
    
    \propag [boson] (w2) to [edge label=$W$] (l2);
    \propag [boson] (l2) to (l1);
    \propag [boson] (l1) to (w2);
    
    \propag [scalar] (l2) to (f2);
    \propag [boson] (l1) to (f1);

    \propag [boson] (w2) to [edge label=$Z/\gamma$] (w1);
\end{feynhand}
\end{tikzpicture}
	\end{flushleft}
	\end{minipage}
		\begin{minipage}{0.3\hsize}
	\begin{flushleft}
			\begin{tikzpicture} 
\begin{feynhand}    
    \vertex [particle] (i1) at (-2,-1.5) {$e^+$};
    \vertex [particle] (i2) at (-2,1.5) {$e^-$};
    \vertex [particle] (f1) at (2,-1.5) {$\gamma$};
    \vertex [particle] (f2) at (2,1.5) {$h$};
    \vertex [particle] (l4) at (-0.5,0.5);
    \vertex [particle] (l3) at (-0.5,-0.5);
    \vertex [particle] (l2) at (0.5,0.5);
    \vertex [particle] (l1) at (0.5,-0.5);
    
    \propag [fermion] (l3) to (i1);
    \propag [fermion] (i2) to (l4);
    
    \propag [boson] (l4) to [edge label=$W$] (l2);
    \propag [boson] (l2) to (l1);
    \propag [boson] (l1) to (l3);
    \propag [fermion] (l4) to (l3);
    
    \propag [scalar] (l2) to (f2);
    \propag [boson] (l1) to (f1);

\end{feynhand}
\end{tikzpicture}
	\end{flushleft}
	\end{minipage}
        \caption{
                \label{fig:sm_feynman} 
              Loop-diagrams in the SM for $e^+e^- \to h \gamma $: (left) top quark loop, (centre) W boson loop, and (right) box diagram with internal $W$ boson lines.}
\end{figure}

\begin{figure}[htbp] 
        
        \centering \includegraphics[width=0.6\columnwidth]{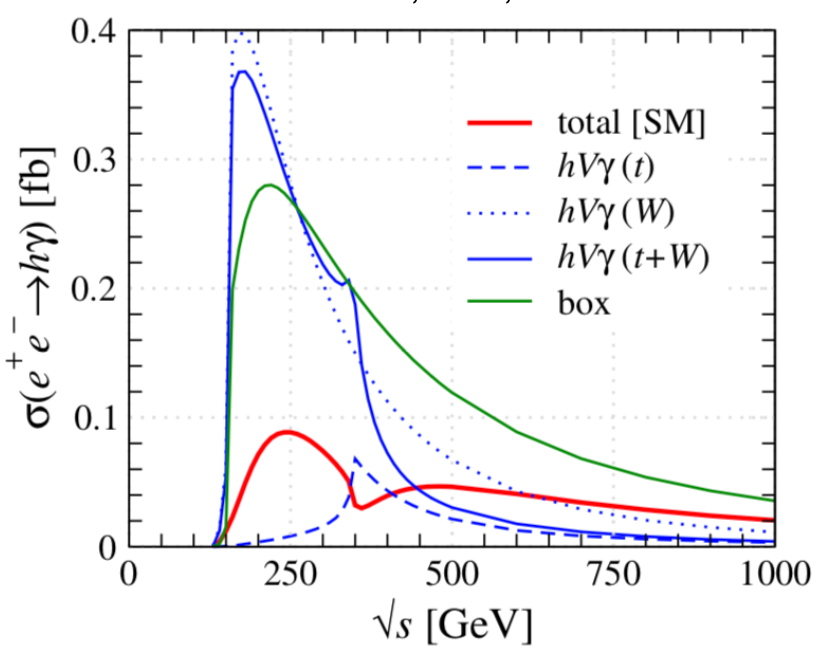}
        \caption{
                \label{fig:crosssection_hg} 
              Contributions from individual diagrams of \Figref{fig:sm_feynman} to $\sigma(e^+e^- \to h \gamma)$. The dashed blue line shows the distribution of the diagram with top-quark in the loop (left), the dotted blue line for the $W$ boson in the loop (middle), the solid blue line shows their sum. The solid green line is for the box diagram with $W$ boson loop (right), and the red line is the total of all contributions~\cite{PhysRevD.99.035023}.}
\end{figure}

The $e^+ e^- \to h \gamma $ process is a loop-induced in the SM. \Figref{fig:sm_feynman} shows the three leading Feynman diagrams for this process in the SM.
\Figref{fig:crosssection_hg} shows the cross section of the $e^+ e^- \to h \gamma $ process as a function of the center of mass energy. The contributions from individual diagrams are shown in the \Figref{fig:crosssection_hg}. We can see that there are significant destructive interferences among different diagrams.

\begin{figure}[htbp] 
\centering
\includegraphics[width=0.8\linewidth]{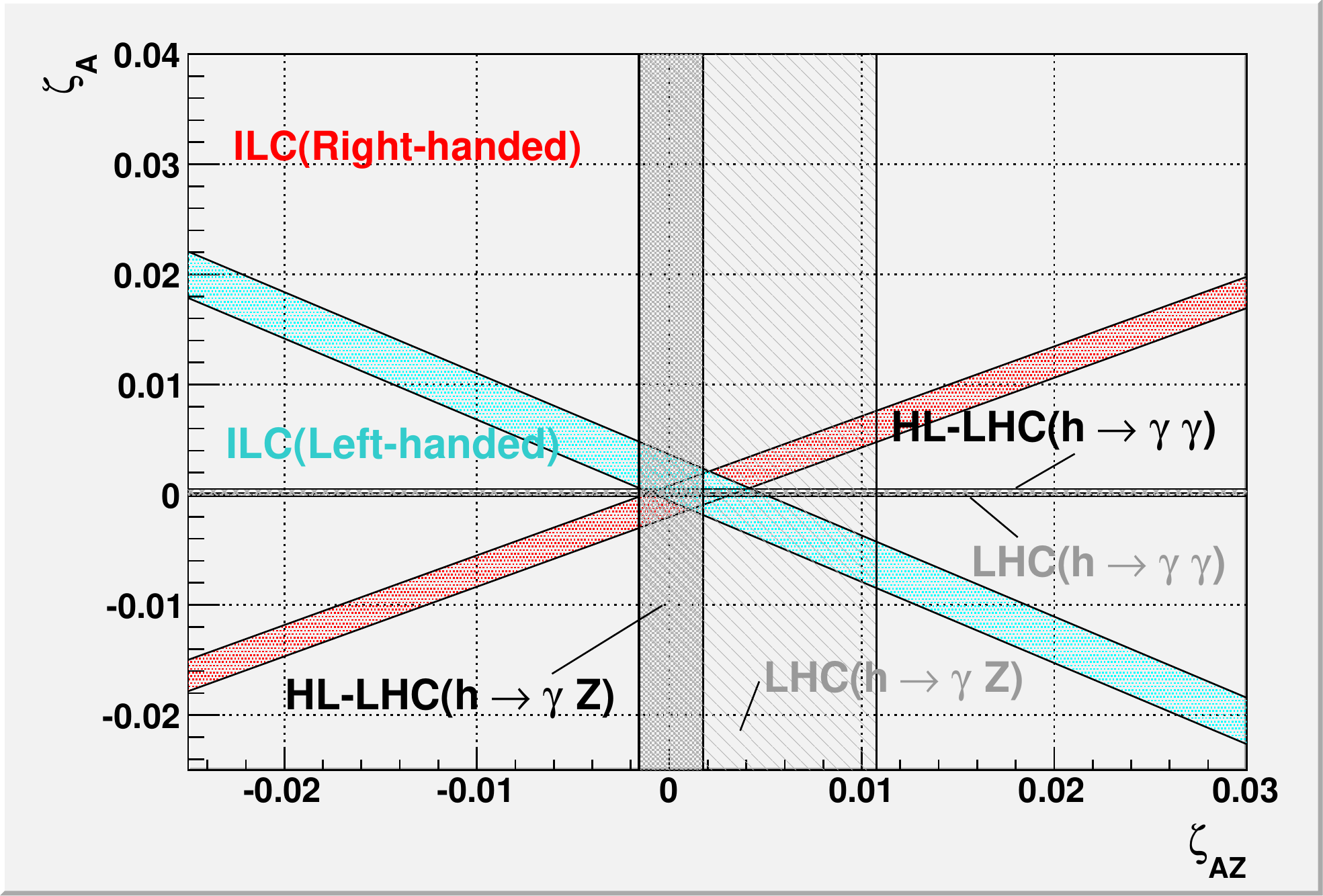}
\caption{
\label{fig:zeta_limit_exp1}
ILC limits on $\zeta_{A}$ and $\zeta_{AZ}$ limitation in the case background-free and perfect selection efficiency, compared to limits from LHC and HL-LHC.
}
\end{figure}

The SM cross sections for four beam polarisation cases are shown in \Tabref{tab:crosssection_hg}. $\sigma_{SM}^L$ represents the cross section for beam polarisation $(P_{e^-}, P_{e^+})=(-100\%,+100\%)$ (left-handed), $\sigma_{SM}^R$ stands for $(P_{e^-}, P_{e^+})=(+100\%,-100\%)$ (right-handed), $\sigma_{SM}^-$ is for $(P_{e^-}, P_{e^+})=(-80\%,+30\%)$, and $\sigma_{SM}^+$ is for $(P_{e^-}, P_{e^+})=(+80\%,-30\%)$. 
These small cross sections for $e^+e^- \to h \gamma$ are a potential advantage, since it could be easier to see the effects of BSM~\cite{PhysRevD.99.035023}\cite{PhysRevD.100.075006}. The expected limits are shown in \Figref{fig:zeta_limit_exp1} in the ideal case background-free and perfect selection efficiency and 100\% signal efficiency using the $h \to b \bar{b}$ channel only.

\begin{table}[htbp]
\caption{SM cross sections for $e^+e^- \to h \gamma $ for different beam polarisations at $\sqrt{s}$  = 250~GeV. $\sigma^{L, R}$ stands for 100\% beam polarisation and $\sigma^{\pm}$ corresponds to 80\% electron and and 30\% positron polarisations.}
\centering
\begin{tabular}{llll} 
\hline
&\multicolumn{1}{c}{$P_{e^-}$ } & \multicolumn{1}{c}{$P_{e^+}$} & \multicolumn{1}{c}{$\sigma_{SM}$[fb]}  \\
\hline
$\sigma_{SM}^L $ & $-100\%$ & $+100\%$ &  0.35\\
$\sigma_{SM}^R $ & $+100\%$ & $-100\%$	& 0.016\\
$\sigma_{SM}^- $ & $-80\%$ & $+30\%$ & 0.20\\
$\sigma_{SM}^+ $ & $+80\%$ & $-30\%$ & 0.021 \\ \hline
\end{tabular}
\label{tab:crosssection_hg} 
\end{table}

\clearpage
\section{Simulation Framework}
First, we generated signal events at 250~GeV with 900~fb$^{-1}$ using \texttt{Physsim}~\cite{physsim} including SM full 1-loop contribution for the matrix element calculation for $(P_{e^-}, P_{e^+}) = (-80\%,+30\%)$ and $(+80\%,-30\%)$ polarisations. Effects of Initial State Radiation(ISR) and beamstrahlung ware included. We took into account all the SM $e^+e^-$ to 2-fermion (2f) and 4-fermion (4f) backgrounds.
We carried out full simulation based on \texttt{Geant4}~\cite{AGOSTINELLI2003250}, with a realistic model of the international large detector using \texttt{Mokka}~\cite{de2002detector}. 
The simulated Monte-Carlo data were fed into a full reconstruction chain from detector signals to reconstructed 4-vectors using Marlin in \texttt{iLCSoft}~\cite{gaede2007marlin}. 
We analysed the $h \to bb$ bar and $h \to WW^*$ semi-leptonic decay modes.

\section{Event Selection and Significance}
The signal has three signatures. First, the signal has one isolated monochromatic photon with an energy of about 93 GeV. Second, it has 2 jets. Third, the invariant mass of the system other than the photon is consistent with the Higgs mass. We focus on the leading Higgs decay channels: $h\to b\bar{b}$ and $h\to WW^*$ (semi-leptonic) final states. 

\subsection{Pre-selection}
To select the signal events, we start with identifying one isolated photon with an energy greater than 50~GeV using $\texttt{PandoraPFA}$. In the case of the $h\to WW^*$ semi-leptonic channel, we require one isolated lepton. The remaining particles are clustered into two jets using the Durham jet clustering algorithm~\cite{CATANI1991432}, and these jets are then flavor tagged using the algorithm in \texttt{LCFI+}~\cite{SUEHARA2016109}. The pre-selection efficiencies for $h\to b\bar{b}$ and $h\to WW^*$ (semi-leptonic) channels are summarised in \Tabref{tbl:sig_bb_left} to \Tabref{tbl:sl_reduction_right}.

\subsection{$h \to b\bar{b}$ Channel}
We require the $b$-likeliness ($b$-likeliness cut) of the two jets greater than 0.77 to suppress events with other flavor jets. 
Total missing energy less than 35~GeV is required because expected missing energy should be small for event without neutrino.

For the remaining background events after these cuts, we performed Multivariate Data Analysis using the Toolkit for Multivariate Data Analysis (TMVA) package~\cite{hoecker2007tmva} of ROOT 5.
We used four input parameters for MVA: the Higgs invariant mass, the photon energy, the angle between the photon and $b$-jets, the angle between the two $b$-jets in the Higgs rest system.
Finally, we required the absolute value of the cosine of the polar angle of the signal photon less than 0.92 since background ISR photons are often in the forward region.
The total number of background, signal, and signal significance, which is given by following formula, for the $h \to b \bar{b}$ channel for $(P_{e^-}, P_{e^+})=(-80\%,+30\%)$ after each event selection are listed in \Tabref{tbl:sig_bb_left}. 
\begin{eqnarray}
\text{Significance}\, (n_{sig}) \equiv \frac{N_{S}}{\sqrt{N_{S}+N_{B}}},
\end{eqnarray}
where $N_{S}$ is the number of selected signal events and $N_{B}$ is the number of selected background events. \Tabref{tbl:sig_bb_right} shows the similar table for $(P_{e^-}, P_{e^+})=(+80\%,-30\%)$. 
After all cuts, we expect 29 signal events and 12 thousand background events for the SM signal process.
In the end, the significance for the Standard Model signal is 0.26$\sigma$ for $(P_{e^-}, P_{e^+})=(-80\%, +30\%)$.
For $(P_{e^-}, P_{e^+})=(+80\%, -30\%)$, only three signal events and $5.9\times10^3$ background event remained with a significance of $3.9 \times 10^{-2}\sigma$.
\begin{table}[htbp]
\centering
\caption{Expected number of events in $h \to b\bar{b}$ channel after applying the selection cut for $(P_{e^-}, P_{e^+})=(-80\%, +30\%)$.}
\label{tbl:sig_bb_left} 
\begin{tabular}{l} $(P_{e^-}, P_{e^+})=(-80\%, +30\%)$, $\int{\mathcal{L}}=900~\mathrm{fb}^{-1}$, $\sqrt{s}=250$~GeV  \end{tabular}
\begin{tabular}
{lccccc}\hline 
 & 2$f$ & 4$f$ & total bg & Signal & Significance \\\hline
Expected & $1.0\times 10^8$ & $3.7\times 10^7$ & $1.4\times 10^8$ & $1.1\times 10^2$ & $9.0\times 10^{-3}$ \\
Pre-selection & $2.8\times 10^7$ & $1.6\times 10^6$ & $2.9\times 10^7$ & $9.9\times 10^1$ & $1.8\times 10^{-2}$ \\
$b$ likeliness$>0.77$ & $2.2\times 10^6$ & $2.1\times 10^4$ & $2.2\times 10^6$ & $9.0\times 10^1$ & $6.0\times 10^{-2}$ \\
$E_{mis}<35$~GeV & $1.9\times 10^6$ & $1.6\times 10^4$ & $1.9\times 10^6$ & $8.2\times 10^1$ & $5.9\times 10^{-2}$ \\
mvabdt $>0.025$ & $1.9\times 10^4$ & $3.2\times 10^2$ & $2.0\times 10^4$ & $3.4\times 10^1$ & $2.4\times 10^{-1}$ \\
-0.92$<cos \theta_{\gamma}<$0.92 & $1.2\times 10^4$& $1.3\times 10^2$ & $1.2\times 10^4$ & $2.9\times 10^1$ & $2.6\times 10^{-1}$ \\ \hline
\end{tabular}
\end{table}
\begin{table}[htbp]
\centering
\caption{Table similar to Table 2 but for $(P_{e^-}, P_{e^+})=(+80\%, -30\%)$.}
\label{tbl:sig_bb_right} 
\begin{tabular}{l} $(P_{e^-}, P_{e^+})=(+80\%, -30\%)$, $\int{\mathcal{L}}=900~\mathrm{fb}^{-1}$, $\sqrt{s}=250$~GeV  \end{tabular}
\begin{tabular}
{lccccc}\hline 
 & 2$f$ & 4$f$ & total bg & Signal & Significance \\ \hline 
Expected & $7.3\times 10^7$ & $4.6\times 10^6$ & $7.8\times 10^7$ & $1.1\times 10^1$ & $1.3\times 10^{-3}$ \\
Pre-selection & $2.3\times 10^7$ & $4.7\times 10^5$ & $2.3\times 10^7$ & $1.0\times 10^1$ & $2.1\times 10^{-3}$ \\
$b$ likeliness$>0.77$ & $1.4\times 10^6$ & $9.3\times 10^3$ & $1.5\times 10^6$ & 9.4  & $7.8\times 10^{-3}$ \\
$E_{mis}<35$~GeV & $1.3\times 10^6$ & $7.7\times 10^3$ & $1.3\times 10^6$ & 8.4  & $7.5\times 10^{-3}$ \\
mvabdt $>$ 0.025 & $1.0\times 10^4$ & $2.1\times 10^2$ & $1.0\times 10^4$ & 3.4  & $3.4\times 10^{-2}$ \\
-0.92 $<cos \theta_{\gamma}<$0.92 & $5.9\times 10^3$ & $5.7\times 10^1$ & $5.9\times 10^3$ & 3.0  & $3.9\times 10^{-2}$ \\ \hline
\end{tabular}
\end{table}
For both $(P_{e^-}, P_{e^+})=(-80\%, +30\%)$ and $(P_{e^-}, P_{e^+})=(+80\%, -30\%)$ beam polarisations, the dominant background comes from continuum region of $e^+e^- \to \gamma b\bar{b}$, which is irreducible.

\subsection{$h\to WW^*$ Semi-leptonic Channel}
To suppress $\mathrm{e}^{+} \mathrm{e}^{-} \to 2l$ background events, we require the number of charged particles in jets to be greater than three.
The $b$-likeness is required to be less than 0.77 to remove overlap with $h \to b \bar{b}$ events. 

A pair of jets or a pair of the lepton and the neutrino as the missing momentum, which has the invariant 
mass closest to $m_W=80.4$ GeV, is combined to form the on-shell $W$, and the rest is defined to originate from the off-shell $W^*$. 
We require $| m_{W_h}-m_W | <10~\mathrm{GeV}$ or $| m_{W_l}-m_W| <9.4~\mathrm{GeV}$, where $m_{W_h}$ is for the hadronic decay, and $m_{W_l}$ for the leptonic decay of the on-shell $W$. 

As with the $b \bar{b}$ channel, the remaining events after these cuts ware fed into multivariate analysis. In this analysis, we used four parameters: the reconstructed Higgs mass, missing energy, the photon energy, and the total visible mass. We required the BDT output value for each event to be greater than 0.1.

\Tabref{tbl:sl_reduction_left} and \Tabref{tbl:sl_reduction_right} show the number of background, signal, and signal significance for $(P_{e^-}, P_{e^+})=(-80\%,+30\%)$ and $(P_{e^-}, P_{e^+})=(+80\%,-30\%)$, respectively. The resulting signal significance is 3.1$\times10^{-1} \sigma$ for $(P_{e^-}, P_{e^+})=(-80\%, +30\%)$, and 4.2$\times10^{-2} \sigma$ for $(P_{e^-}, P_{e^+})=(+80\%, -30\%)$.

\begin{table}[htbp]
\centering
\caption{Expected number of events in the $h \to WW^*$ semi-leptonic channel for $(P_{e^-}, P_{e^+})=(-80\%, +30\%)$, after applying each selection cut.}
\label{tbl:sl_reduction_left} 
\begin{tabular}{l} $(P_{e^-}, P_{e^+})=(-80\%, +30\%)$, $\int{\mathcal{L}}=900~\mathrm{fb}^{-1}$, $\sqrt{s}=250~GeV$  \end{tabular}
\begin{tabularx}{\linewidth}{p{45mm} XXXXX} \hline
 & 2$f$ & 4$f$ & total bg & Signal & Significance \\ \hline 
Expected & $1.0\times 10^8$ & $3.7\times 10^7$ & $1.4 \times 10^8$ & $1.8\times 10^1$ & $3.4\times 10^{-3}$ \\
Pre-selection & $1.3\times 10^7$ & $7.5\times 10^5$ & $1.3 \times 10^7$ & $1.0\times 10^1$ & $3.6\times 10^{-3}$ \\
\# of charged particle$>$3 & $7.8\times 10^4$ & $2.3\times 10^5$ & $3.1 \times 10^5$ & 5.4 & $9.8\times 10^{-3}$ \\
$| m_{W_h}-80.4 | <10~\mathrm{GeV}$ \& $| m_{W_l}-80.4| <9.4~\mathrm{GeV}$  & $2.5\times 10^4$ & $1.6\times 10^5$ & $1.9 \times 10^5$ & 3.7 & $8.6\times 10^{-3}$ \\
$b$ likeliness$<0.77$ & $1.7\times 10^4$ & $1.6\times 10^5$ &$ 1.8 \times 10^5$ & 3.7 & $8.7\times 10^{-3}$ \\
mvabdt $>$ 0.1 & 3.1 & $3.8\times 10^1$ & $4.1\times 10^1$ & 1.0 & $1.6\times 10^{-1}$ \\
-0.93$<cos \theta_{\gamma}<$0.93 & 0.0 & 8.4 & 8.4 & $9.5\times 10^{-1}$ & $3.1\times 10^{-1}$ \\  \hline
\end{tabularx}
\end{table}

\begin{table}[htbp]
\centering
\caption{Table similar to Table 4, but for $(P_{e^-}, P_{e^+})=(+80\%, -30\%)$.}
\label{tbl:sl_reduction_right} 
\begin{tabular}{l} $(P_{e^-}, P_{e^+})=(+80\%, -30\%)$, $\int{\mathcal{L}}=900~\mathrm{fb}^{-1}$, $\sqrt{s}=250$~GeV  \end{tabular}
\begin{tabularx}{\linewidth}{p{45mm} XXXXX} \hline
 & 2$f$ & 4$f$ & total bg & Signal & Significance \\ \hline
Expected & $7.3\times 10^7$ & $4.6\times 10^6$ & $7.8 \times 10^7$ & 1.9 & $4.8\times 10^{-4}$ \\
Pre-selection & $1.2\times 10^7$ & $3.1\times 10^5$ & $1.2 \times 10^7$ & 2.0& $ 3.9\times 10^{-4}$ \\
\# of charged particle $>$ 3 & $5.0\times 10^4$ & $3.6\times 10^4$ & $8.6 \times 10^4$ & 1.5 & $1.9\times 10^{-3}$ \\
$| m_{W_h}-80.4 | <10~\mathrm{GeV}$ \& $| m_{W_l}-80.4| <9.4~\mathrm{GeV}$  & $1.7\times 10^4$ & $1.5\times 10^4$ & $3.2 \times 10^4$ & $3.8\times 10^{-1}$ & $2.1\times 10^{-3}$ \\
$b$ likeliness$<0.77$ & $1.2\times 10^4$ & $1.4\times 10^4$ & $2.6 \times 10^5$ & $3.7\times 10^{-1}$ & $2.3\times 10^{-3}$ \\
mvabdt $>$ 0.1 & $5.3\times 10^1$ & $2.1\times 10^1$ & $7.4\times 10^1$ & $1.0\times 10^{-1}$ & $1.2\times 10^{-2}$ \\
-0.93$<\cos \theta_{\gamma}<$ 0.93 & 0.0 & 4.7 & 4.7 & $9.3\times 10^{-2}$ & $4.2\times 10^{-2}$ \\ \hline
\end{tabularx}
\end{table}

\clearpage
\section{Result: 95 $\%$ Confidence Level Upper Limit for the Cross Section of $e^+e^- \to h \gamma $}
Since our full simulation study showed that the $e^+e^- \to h\gamma$ process is difficult to observe for both $h \to b \bar{b}$ and $h \to WW^*$ decay modes, we put 95 $\%$ Confidence Level upper limits on the cross section using \Equref{upperlimit_calc}. \Tabref{upperlimit} is a summary of 95 $\%$ Confidence Level upper limits on the cross section for two decay modes and two polarization cases:$(P_{e^-}, P_{e^+})=(-80\%,+30\%)$ and $(P_{e^-}, P_{e^+})=(+80\%,-30\%)$.
\begin{eqnarray}
\label{upperlimit_calc} 
\sigma_{h \gamma}=\sigma_{SM}+\frac{1.645}{\text { significance }} \sigma_{SM}
\end{eqnarray}

\begin{table}[htbp]
\centering
\caption{Summary of the 95 $\%$ confidence level upper limit on the cross section of $e^+e^- \to h \gamma $.}
\label{upperlimit} 
\begin{tabular}
{lcc}\hline 
\text { Channel } & \text { Beam polarisation$(P_{e^-}, P_{e^+})$ } & \text {95\% C.L. Upper limit [fb]} \\ 
\hline 
 \text { $h \to b\bar{b}$ } & $(-80\%, +30\%)$  & $\sigma_{h\gamma}^-<1.5$  \rule[0mm]{0mm}{5mm} \\ 
\text { $h \to b\bar{b}$ } & $(+80\%, -30\%)$ & $\sigma_{h\gamma}^+<0.9$\\ 
\text { $h \to WW^*$ semi-leptonic } &  $(-80\%, +30\%)$ & $\sigma_{h\gamma}^-<1.3$ \\
\text { $h \to WW^*$ semi-leptonic }&  $(+80\%, -30\%)$ & $\sigma_{h\gamma}^+<0.8$ \\ \hline
\end{tabular}

\end{table}
We combined the results of the two signal channels $h\to b \bar{b}$ and $h\to WW^*$, for each polarisation by calculating the square root of the sum of squares of significance. We then calculated the combined 95$\%$ C.L. upper limits on the signal cross sections. The results in the unit of the SM cross section are as follows:
\begin{eqnarray}
\frac{\sigma_{h \gamma}^-}{\sigma_{S M}^-} < 5.1 ~(\text{95\% C.L.})\\
\frac{\sigma_{h \gamma}^+}{\sigma_{S M}^+} < 28.3~(\text{95\% C.L.}),
\label{Equ:11}
\end{eqnarray}
where we used $\sigma_{S M}^{-}=0.20~\mathrm{fb}$, and $\sigma_{SM}^{+}=0.021~\mathrm{fb}$.

\section{Constraint on $h\gamma Z$ Coupling}
We convert the cross section for $(P_{e^-}, P_{e^+})=(-80\%,+30\%)$ and $(P_{e^-}, P_{e^+})=(+80\%,-30\%)$ to the 100\% polarization case to constrain anomalous $h \gamma Z$ coupling, $\zeta_{AZ}$.

The cross sections for the finite polarisations (namely $\sigma_{h\gamma}^{-}$ and $\sigma_{h\gamma}^{+}$) and those for the 100\% polarisations(namely $\sigma_{h\gamma}^{L}$ and $\sigma_{h\gamma}^{R}$) are related by the following equation:
\begin{eqnarray}
\left(\begin{array}{l}\sigma_{h\gamma}^{-} \\ \sigma_{h\gamma}^{+}\end{array}\right)=\left(\begin{array}{ll}\frac{1}{4}\left(1+|P_{e^{-}}|\right)\left(1+|P_{e^{+}}|\right) & \frac{1}{4}\left(1-|P_{e^-}|\right)\left(1-|P_{e^+}|\right) \\
\frac{1}{4}\left(1-|P_{e^-}|\right)\left(1-|P_{e^+}|\right) & \frac{1}{4}\left(1+|P_{e^-}|\right)\left(1+|P_{e^+}|\right)\end{array}\right)\left(\begin{array}{l}\sigma_{h\gamma}^{L} \\ \sigma_{h\gamma}^{R}\end{array}\right)
\end{eqnarray}
where $(|P_{e^-}|, |P_{e^+}|)=(80\%,\,30\%)$ in the ILC case. 
The statistical errors on $\sigma_{h\gamma}^L$ and $\sigma_{h\gamma}^R$ are $\Delta \sigma_{h\gamma}^{L} = 0.85~\mathrm{fb}$ and $\Delta \sigma_{h\gamma}^{R} = 0.60~\mathrm{fb}$, respectively.

We can then calculate the significance for $(P_{e^-},\,P_{e^+})=(-100\%,\,+100\%)$, $n_{\text {sig}}^{L}$, and that for $(P_{e^-},\,P_{e^+})=(+100\%,\,-100\%)$, $n_{\text {sig }}^{R}$ as follows: 
\begin{eqnarray}
n_{\text {sig }}^{L}&=&\frac{\sigma_{SM}^L}{\Delta \sigma_{h\gamma}^{L}} =\frac{0.35~\mathrm{fb}}{0.85~\mathrm{fb}}=0.41 \\ \rule{0in}{6ex}
n_{\text {sig }}^{R}&=&\frac{\sigma_{SM}^R}{\Delta \sigma_{h\gamma}^{R}}
=\frac{0.0 16~\mathrm{fb}}{0.60~\mathrm{fb}}=0.027 .
\end{eqnarray}
Finally the 95\% confidence level upper limits for the 100\% polarisations in units of the SM cross section are
\begin{eqnarray}
\frac{\sigma_{h \gamma}^L}{\sigma_{SM}^L} &<& 5.0 ~(\text{95\% C.L.})\\
\frac{\sigma_{h\gamma}^R}{\sigma_{SM}^R} &<& 61.9 ~(\text{95\% C.L.}).
\end{eqnarray}

Using these upper limits on the cross sections, we evaluated the limit on the SMEFT coefficient, $\zeta_{AZ}$. 
Relations between the cross section and the coefficients are given by following formulae~\cite{PhysRevD.94.095015}:
\begin{eqnarray}
\frac { \sigma _ { h\gamma  }^L } { \sigma _{ S M }  ^{L}} = 1 - 201 \zeta _ { A } - 273 \zeta _ { A Z }
\label{Equ:leftcrosssection} \\
\frac { \sigma _ { h \gamma  }^R } { \sigma _ { S M }^R } = 1 + 492 \zeta _ { A } - 311 \zeta _ { A Z }.
\label{Equ:rightcrosssection}
\end{eqnarray}

Assuming $\zeta_{A}$ is 0 and $\sigma_{h\gamma}>0$, we can constrain $\zeta_{AZ}$ for each polarisation. 
For the left-handed case, the constraint on the $\zeta_{AZ}$ is
\begin{eqnarray}
\text { 5.0 } > \quad \frac{\sigma_{h \gamma}^L}{\sigma_{S M}^L}=1-273 \zeta_{A}-201 \zeta_{A Z} >0\\
-0.020< \zeta_{A Z}<0.005,
\end{eqnarray}
For the right handed case the constraint is:
\begin{eqnarray}
 61.9>\frac{\sigma_{h \gamma}^R}{\sigma_{S M}^R}=1+492 \zeta_{A}-311 \zeta_{A Z}>0\\
-0.195<\zeta_{A Z}<0.003.
\end{eqnarray}
The combined $\zeta_{AZ}$ limit is $-0.020<\zeta_{A Z}<0.003$.

We compare our limits with the current LHC results and the HL-LHC projections as shown in \Tabref{tb:lhc_results}.

The LHC limit on $h \to \gamma \gamma$ is taken from the ATLAS experiment~\cite{ATLASCONFNote}, and the LHC limit on $h \to \gamma Z$ from the CMS experiment~\cite{CMS-PAS-HIG-19-014}. These results provide 1$\sigma$ limits for $\sigma \times BR(h \to \gamma \gamma)$ of 127 $\pm$ 10~fb and $\frac{BR(h \to Z \gamma)}{BR(h \to \gamma \gamma)}$ of 1.54$^{+0.65}_{-0.58}$, thus we calculate the upper and lower values of these limits divided by its SM value $(\sigma \times BR(h \to \gamma \gamma))_{SM}=116\pm$ 5~fb, and $\frac{BR(h \to Z \gamma)}{{BR(h \to \gamma \gamma)}}_{SM}=0.69\pm 0.04$. Observed values contain their SM values, therefore we should subtract one from observed values.
\begin{eqnarray}
\frac{(\sigma \times BR)_{obs}}{(\sigma \times BR)_{SM}} &=& \frac{(\sigma \times BR)_{SM}+(\sigma \times BR)_{BSM}}{(\sigma \times BR)_{SM}}\\ \nonumber
&=& 1+ \frac{BR_{BSM}}{BR_{SM}}. 
\end{eqnarray}

We convert the HL-LHC projections from ATLAS to the limits $-0.072<526 \zeta_{A}<0.076$ and $-0.46<290 \zeta_{\mathrm{AZ}}<0.51$.

\begin{table}[htbp]
\centering
\caption{Current LHC limits and HL-LHC projections for SMEFT coefficients $\zeta_{AZ}$ and $\zeta_{A}$. $\zeta_{A}$ and $\zeta_{AZ}$ coefficients are defined in the Introduction.}
\label{tb:lhc_results} 
\begin{tabular}{lc}\hline 
Process & Limitation\\ \hline
LHC limit ($h \to \gamma \gamma$) (Measured) & $ 0.92<1+526 \zeta_{\mathrm{A}}<1.27$~\cite{ATLASCONFNote}\\
LHC limit ($h \to \gamma Z$) (Measured) & $0.55<1+290 \zeta_{A Z}<4.12$~\cite{CMS-PAS-HIG-19-014}\\
HL-LHC limit ($h \to \gamma \gamma$) (Expected) & $-0.072<526 \zeta_{A}<0.076$~\cite{ATLAS:2018jlh}\\
HL-LHC limit ($h \to \gamma Z$) (Expected) & $-0.46<290 \zeta_{\mathrm{AZ}}<0.51$~\cite{ATLAS:2018jlh}\\ \hline
\end{tabular}
\end{table}

\begin{figure}[htbp] 
\centering
\includegraphics[width=0.8\linewidth]{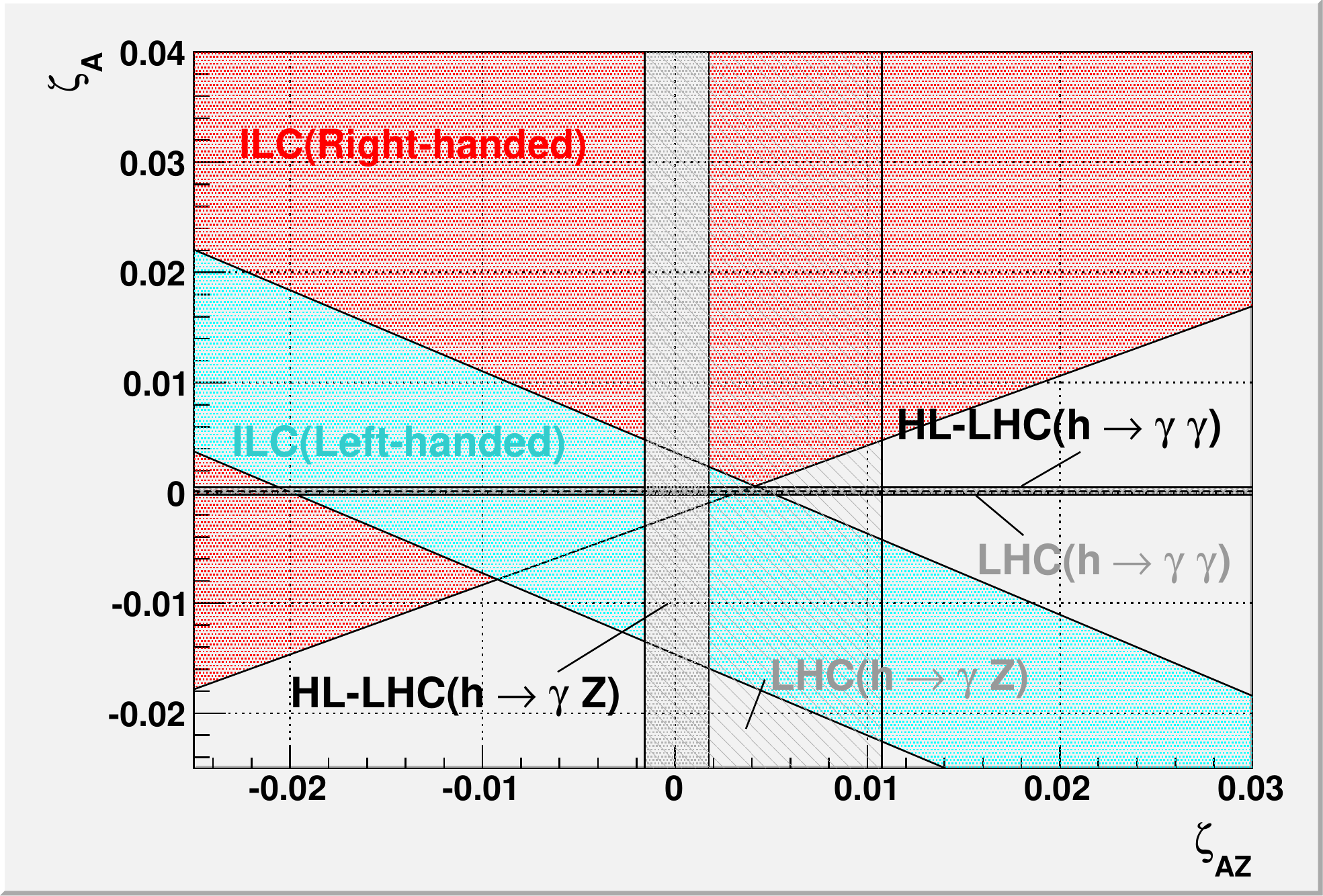}
\caption{
\label{fig:zeta_limit}
Current projected limits on $\zeta_{A}$ and $\zeta_{AZ}$. 
The blue and red areas are our left- and right-handed limits at $\sqrt{s}=250$~GeV ILC, respectively. The grey area shows the current limits from the LHC, and dark grey area shows the HL-LHC projection.
}
\end{figure}

These results are shown in~\Figref{fig:zeta_limit}. 
Our results, which are indicated by blue and red areas turned out to be not strong enough to constrain $\zeta_{A}$ and $\zeta_{AZ}$ beyond the expected limits from HL-LHC, shown by grey areas.

\clearpage
\section{Conclusion}
We performed a full simulation study of the $h \gamma Z$ measurement using $e^+e^- \to h \gamma$ process at the ILC. 
We used a realistic ILD detector model and 250 GeV ILC with 900 fb$^{-1}$ integrated luminosity for each of $(P_{e^-},P_{e^+})$~$=(-80\%,+30\%)$ and $(P_{e^-},P_{e^+})=(+80\%,-30\%)$ polarisations. The reconstructed events were analysed in $h \to b\bar{b}$, $WW^*$ semi-leptonic decay modes.
The signal significance for both $h \to b \bar{b}$ and $h \to WW^*$ semi-leptonic channel turned out to be much less than 1$\sigma$. We, hence, put the upper limit on the signal cross section by combining the results for the $b\bar{b}$ and $WW^*$ semi-leptonic channels for purely left- and right-handed polarisations. 
The resultant 95\% C.L. upper limits on the $h \gamma$ production cross section for the SM are found to be $\frac{\sigma_{h \gamma}^{L}}{\sigma_{S M}^{L}}<5.0$ and $\frac{\sigma_{h \gamma}^{R}}{\sigma_{S M}^{R}}<61.9$ for 100\% left- and right-handed polarisations, respectively. The upper limits constrain the EFT coefficient $\zeta_{AZ}$ as -0.020$<\zeta_{A Z}<$0.003. Comparing with current LHC result and HL-LHC projection, our expected limits turned out to be rather weak.

\section*{Acknowledgements}
We would like to thank the LCC generator working group and the ILD software working group for providing the simulation and reconstruction tools and producing the Monte Carlo samples used in this study.
This work has benefited from computing services provided by the ILC Virtual Organization, supported by the national resource providers of the EGI Federation and the Open Science GRID.
This work is supported in part by the Japan Society for the Promotion of Science under the Grants-in-Aid for Science Research 16H02173.

\clearpage
\bibliographystyle{unsrt}
\bibliography{ref}

\end{document}